\documentclass[final,3p,twocolumn]{elsarticle}

\usepackage[numbers]{natbib}
\usepackage[cal=cm]{mathalfa}
\usepackage{amsmath, amssymb, mathtools} 
\usepackage{graphicx}
\usepackage{siunitx}
\usepackage{float}
\usepackage{url}
\usepackage{color, xcolor}
\usepackage{tikz}
\usepackage{balance}
\usepackage{hyperref}
\usepackage{cleveref}
\crefname{lstlisting}{listing}{listings}
\usepackage{xspace}
\usepackage{comment}
\usepackage{etoolbox}
\usepackage{amsthm}
\usepackage{tabularx, booktabs, array}
\newcolumntype{C}{>{\centering\arraybackslash}X}

\usepackage{algorithm}
\usepackage{algpseudocode}

\usepackage{listings}
\lstdefinelanguage{JSON}{
  morestring=[b]",
  morecomment=[l]{//},
  morekeywords={true,false,null},
  sensitive=false,
}
\usepackage{soul}         
\usepackage{eucal}        
\usepackage{minibox}      

\newcommand{\DisCTI}{\texttt{DisCTI}\xspace}

\algnewcommand{\Inputs}[1]{%
  \Statex \textbf{Inputs:}
  \Statex \hspace*{\algorithmicindent}\parbox[t]{.8\linewidth}{\raggedright #1}
}
\algnewcommand{\Outputs}[1]{%
  \Statex \textbf{Outputs:}
  \Statex \hspace*{\algorithmicindent}\parbox[t]{.8\linewidth}{\raggedright #1}
}
\algnewcommand{\Initialize}[1]{%
  \Statex \textbf{Initialize:}
  \Statex \hspace*{\algorithmicindent}\parbox[t]{.8\linewidth}{\raggedright #1}
}

\newcommand{\subheading}[1]{
    \vspace{.6pt}
    \noindent{\textit{\textbf{#1.}}}
}

\def\tsc#1{\csdef{#1}{\textsc{\lowercase{#1}}\xspace}}
\tsc{WGM}
\tsc{QE}
\tsc{EP}
\tsc{PMS}
\tsc{BEC}
\tsc{DE}

\begin{document}
\begin{frontmatter}

\let\WriteBookmarks\relax
\def\floatpagepagefraction{1}
\def\textpagefraction{.001}


\title{\DisCTI: Who Needs to Know Timely? Automated Sector-Aware Cyber Threat Intelligence Dissemination}



\author[1]{Fajar Wijitrisnanto} 

\author[2]{Alsharif Abuadbba} 

\author[2,3]{Yansong Gao} 

\author[2]{Nan Wu}

\affiliation[1]{organization={National Cyber and Crypto Agency},
                city={Jakarta},
                country={Indonesia}}

\affiliation[2]{organization={CSIRO},
                city={Sydney},
                country={Australia}}

\affiliation[3]{organization={The University of Western Australia},
                city={Perth},
                country={Australia}}


\begin{abstract}
The timely dissemination of cyber threat intelligence (CTI) is critical for organizations to mount swift and effective incident response. When valid CTI is delivered to the right sector at the right time, identical attacks can often be contained or mitigated. However, today's rapidly expanding CTI landscape overwhelms analysts, who must sift through massive and heterogeneous feeds. Existing platforms such as the Malware Information Sharing Platform (MISP) provide sector tagging features (e.g., energy, finance, government), but in practice, these remain largely unmapped—98\% of events are left uncategorized. This lack of automated and timely sector mapping severely limits the operational value of shared intelligence, leaving organizations that belong especially to the critical information infrastructure sector exposed.

To address this gap, we formulate sector-targeted CTI dissemination as a multilabel classification problem. Leveraging deep field knowledge of CTI structures and sector-specific threat patterns, we construct a novel data set of 872 sector-labelled CTI events from a threat intelligence platform (TIP). We then apply BERT, a transformer-based model, to automate the mapping of CTI events to sectors. Using the structured threat information expression (STIX) format for cross-platform interoperability, our approach achieves a macro-averaged F1-score of 0.89 at a Hamming loss of 0.055 on the custom dataset, i.e.\ 94.5\% of individual sector-label assignments are correct. These results not only demonstrate the feasibility of sector-aware, automated CTI dissemination but also highlight how embedding expert field knowledge into machine learning design fills a crucial gap in the threat intelligence pipeline, enabling faster and context-relevant defensive action.
\end{abstract}



\end{frontmatter}

\section{Introduction}
The cybersecurity landscape is rapidly escalating in scale and sophistication, with both national and global reports highlighting alarming trends~\cite{wong2022phishclone,wang2023doitrust,chen2024peek,abuadbba2026promise}. The 2024 security monitoring report from the Indonesian National Cyber and Crypto Agency (BSSN) recorded over 330 million network anomalies, more than 2.4 million Advanced Persistent Threat (APT) activities linked to groups such as Lazarus, Wicked Panda, and APT 40, and over half a million ransomware incidents in a single year~\cite{BSSN2024Landscape}. These figures illustrate not only the breadth but also the increasing sophistication of malicious activities. A similar surge is evident globally: ENISA’s Threat Landscape 2023 report documented a doubling of cybersecurity incidents from 2022 to 2023~\cite{enisathreat2023}, while IBM’s Cost of a Data Breach 2024 reported the highest average breach cost to date~\cite{ibmcost2024}. Together, these statistics underscore the pressing need for organizations to strengthen their cybersecurity posture with timely, sector-aware measures capable of detecting, responding to, and mitigating threats at speed.

In response to escalating cyber threats, cyber threat intelligence (CTI)  has emerged as a cornerstone of modern defense strategies. CTI refers to the systematic collection, analysis, and dissemination of information on potential or ongoing threats~\cite{technicalcti2018}. Its ultimate goal is to prepare organizations to prevent attacks where possible and to enable timely and effective incident response when prevention fails.

\begin{figure*}[t]
\centering
\includegraphics[width=0.95\textwidth,trim={0 0 0 0cm},clip]{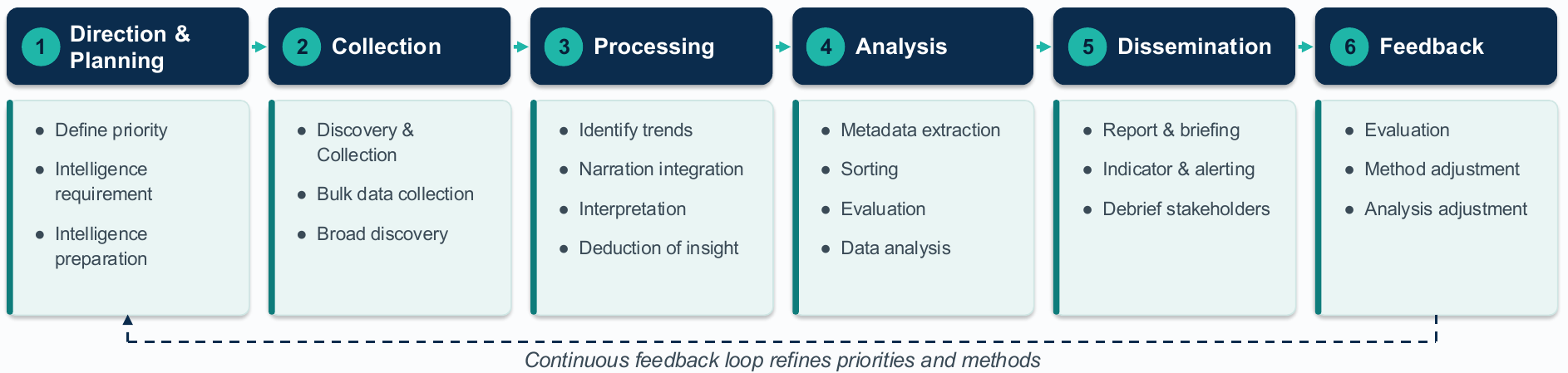}
\caption{The Cyber Threat Intelligence Life-Cycle}
\label{fig:0}
\end{figure*}

The implementation of CTI commonly follows frameworks such as the Threat Intelligence Life Cycle illustrated in Figure~\ref{fig:0}. The cycle consists of six interdependent stages. (1) Direction. In this first stage, organizations define their intelligence objectives and requirements, which shape all subsequent CTI activities. (2) Collection. The raw data are then collected from various internal and external sources such as CTI feeds, network logs, application logs, and open-source intelligence (OSINT). (3) Processing. The collected data is then cleaned, normalized, and enriched to transform it into a structured format suitable for analysis. (4) Analysis. At this stage, the processed data is examined to identify trends, patterns, or anomalies that indicate potential threats, turning raw information into actionable insights. (5) Dissemination. These insights are distributed to the relevant stakeholders, ensuring that the right information reaches the right people at the right time. (6) Feedback. Finally, stakeholders provide input on the accuracy, usefulness, and timeliness of the intelligence, which is used to refine and improve the entire cycle.

While all stages of the CTI life-cycle are important, the dissemination phase is particularly critical. Intelligence that is not delivered promptly and contextualized for the right sector or stakeholder quickly loses its operational value. \textit{For critical infrastructure operators—such as those in energy, finance, and healthcare—one of the most pressing challenges is knowing, in a timely manner, which attacks specifically target their sector}~\cite{cisaCriticalInfra}. Unlike generic CTI sharing, effective dissemination ensures that actionable intelligence reaches the right entity at the right time, enabling swift and targeted defensive measures. This makes timely, sector-aware dissemination central to transforming CTI from raw data into meaningful and practical defense.

CTI sharing broadly exchanges threat information across organizations and communities to enhance collective cybersecurity. In contrast, CTI dissemination is targeted and actionable, focusing on delivering the right intelligence to the right stakeholders at the right time. This precision is crucial, as organizations—especially in critical sectors—can respond swiftly only if attacks targeting them are promptly identified. However, the rapid growth of threat data makes timely dissemination a major challenge. Effective dissemination solutions are therefore essential to ensure that valid CTI reaches its intended recipients, enabling faster mitigation of identical threats and strengthening overall security posture~\cite{Wagner2016MISP}.

\begin{figure}[htbp]
    \centering
    \includegraphics[width=0.50\textwidth]{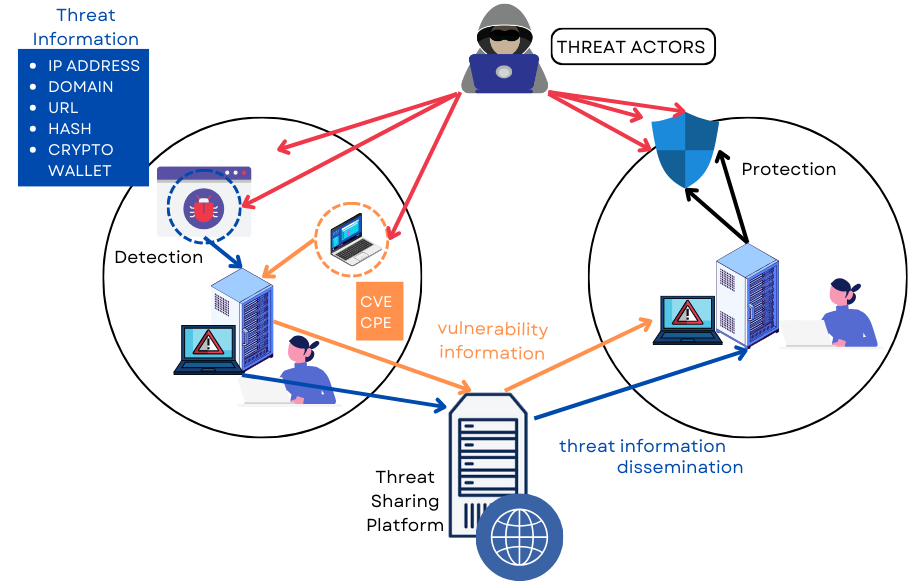}
    \caption{Cyber threat information dissemination}
    \label{fig:3}
\end{figure}

To address CTI dissemination challenges, many sectors rely on information sharing and analysis centers (ISACs)—trusted, non-profit communities that facilitate cyber threat information exchange between public and private stakeholders (ENISA Report: ISAC Cooperative Models). ISACs foster collaboration through meetings, working groups, and joint research, circulating diverse intelligence such as indicators of compromise (IoCs), TTPs, alerts, reports, and configuration rules, as illustrated in \Cref{fig:3}. While these efforts draw on varied sources (e.g., threat feeds, internal logs, and sharing communities), the manual analysis required to contextualize and route intelligence remains resource-intensive. This often results in alert fatigue, where the sheer volume of data overwhelms analysts and undermines timely, effective decision-making~\cite{Tariq2025AlertFatigue}. Overcoming this bottleneck is critical to realizing the operational value of CTI.

Several threat intelligence platforms such as AlienVault open threat exchange (OTX), OpenCTI, and MISP, provide automation features to streamline CTI collection, analysis, and sharing. Yet, a key limitation remains: most CTI data—particularly from open sources—lacks sector-specific tagging, which is critical for precision dissemination. Without it, organizations are forced to sift through vast amounts of irrelevant intelligence rather than focusing on threats directly impacting their sector.

Since CTI events are often relevant to multiple sectors simultaneously, this challenge is best formulated as a multilabel classification problem~\cite{Chen2022}. For instance, a single threat report may apply to both the Finance and Healthcare sectors. \textit{Addressing this requires advanced machine learning models capable of accurate multilabel tagging, enabling targeted dissemination that delivers actionable intelligence to the right sectors without overwhelming them with noise, enabling such a solution to unlock the full operational value of CTI dissemination, and to the best of our knowledge, this is an open and unresolved challenge.}\\

\textbf{Contributions.} In this paper, we addressed the above challenges, and the key contributions are as follows:

\begin{itemize}
    \item \textit{DisCTI} Framework. We design \textit{DisCTI}, a framework for automated sector-specific CTI dissemination, leveraging multi-label classification.
    \item Novel Dataset. We construct the first sector-labeled CTI dataset by collecting and processing 872 real-world events from a Threat Intelligence Platform, enabling systematic research on sector-aware dissemination.
    \item Comparative Evaluation. We design and evaluate three multilabel classification approaches—Parallel Binary Classifiers, Sequential Binary Classifiers, and a transformer-based BERT model—and identify the best-performing approach for the sector-tagging task.
    \item Empirical Validation. Our experiments show that the BERT model outperforms the baselines, achieving a macro-averaged F1-score of 0.89 at a Hamming loss of 0.055, establishing a strong benchmark for automated, sector-specific CTI dissemination.
\end{itemize}

By demonstrating the feasibility of automated sector tagging, our work advances the operational value of CTI dissemination and supports faster, more targeted defensive responses in critical infrastructure environments.

The remainder of this paper is structured as follows. Section 2 reviews preliminary concepts and related work in Cyber Threat Intelligence. Section 3 outlines our methodology for data collection and the construction of the sector-labeled dataset. Section 4 introduces our automated tagging models, including both Binary Classifier approaches and the BERT-based classifier. Section 5 reports and compares the experimental results. Section 6 concludes with key findings and highlights directions for future research.

\section{Preliminary \& Related Work}\label{sec:related work}
In this section, we review existing research on CTI standards, platforms, and dissemination approaches. We first describe how structured formats such as STIX enable interoperability, then discuss how threat intelligence platforms aggregate and analyze intelligence. Finally, we highlight dissemination challenges, particularly the lack of sector-specific automation, and position our work as a multi-label learning approach to address this gap.

\subsection{Threat Intelligence Data with STIX}
STIX is a standardized language and serialization format used to convey cyber threat intelligence \cite{OASIS_STIX}. It facilitates the sharing and dissemination of CTI across organizational boundaries, enhancing situational awareness and improving response capabilities~\cite{STIXmitre2012}.

STIX includes a range of data elements called STIX Data Object (SDO) such as indicators, observables, threat actors, attack patterns, and campaigns, enabling detailed representation of cyber threats. This structured approach allows for the consistent description of threat information, making it easier to automate and integrate into threat intelligence platforms.

For example, a STIX report might describe a Threat Actor object (like a known hacking group) who is observed using a specific Attack Pattern object (like spearphishing). This attack could be linked to an Indicator object (a malicious URL) and target an Identity object (a company in the finance sector). These objects are connected through STIX Relationship Objects (SROs), creating a comprehensive and machine-readable graph of the threat.

The adoption of STIX as a standard for threat intelligence sharing has been widely supported by organizations such as the Cyber Threat Intelligence Integration Center (CTIIC) and the European Union Agency for Cybersecurity (ENISA). Research indicates that using STIX for threat information sharing improves the accuracy and efficiency of threat detection and response processes~\cite{jin2024ctisharing}.

\subsection{Threat Intelligence Platform} 

TIPs are designed to aggregate, analyze, and share threat intelligence data from multiple sources to provide actionable insights to security teams. TIPs serve as centralized repositories for threat data, integrating various data sources, including security logs, threat feeds, and open-source intelligence.

TIPs like AlienVault OTX \cite{Otx}, ThreatConnect \cite{ThreatConnect}, and IBM X-Force Exchange \cite{IbmXForce} are equipped to aggregate vast amounts of threat data from diverse sources. This aggregation is critical as it enables the collection of comprehensive threat information that spans various domains and geographies~\cite{Sauerwein2017ThreatIS}. These platforms support the integration of structured threat intelligence formats such as STIX and TAXII (Trusted Automated eXchange of Indicator Information), ensuring that threat data is not only collected but also shared and consumed in a standardized manner \cite{OASIS_TAXII}. This integration facilitates automated and streamlined data exchange processes between different entities and systems~\cite{TAXIImitre2012}.  

One of the core functionalities of TIPs is their ability to analyze and correlate threat data to identify patterns, trends, and anomalies. This involves advanced analytics, including machine learning algorithms, to detect potential threats and provide contextual insights. Another functionality of TIPs is related to CTI visualization. Platforms like OpenCTI provide sophisticated visualization tools that enable security teams to map out threat actor activities, attack vectors, and the relationships between different threat entities. This helps in creating a comprehensive threat landscape view, facilitating better decision-making.

However, a significant challenge remains that these platforms do not inherently address. While TIPs excel at aggregating and visualizing threat data, \textit{they generally lack sophisticated, built-in mechanisms for automated, sector-specific tagging and dissemination}. The contextualization of a threat—determining if it is relevant to the 'finance', 'health', or 'energy' sector—is a critical step that is still largely a manual, labor-intensive process for security analysts. This forces teams to sift through vast amounts of aggregated data to find what is relevant, directly leading to the problem of information fatigue and delayed responses. \textit{It is this specific gap—the lack of automated, granular, sector-based tagging at scale—that our work aims to solve.}

\subsection{Threat Intelligence Dissemination}
The increasing volume of CTI data poses a significant challenge for security analysts who need to process and analyze extensive CTI feeds. Targeted CTI dissemination solutions are necessary to streamline this process, allowing organizations to concentrate on the most pertinent information~\cite{CIRCL2020Statistic}. Leveraging automated TIPs is crucial for logging and tagging security incidents or events, which ensures contextual relevance.

Despite their capabilities, TIPs face challenges related to the contextual relevance of the intelligence they provide. The effectiveness of TIPs is heavily dependent on the accurate tagging and classification of threat data, which is often underutilized. For instance, in the CIRCL MISP platform for private sector (MISPPriv), only about 2\% of security events are tagged according to the organization sector, highlighting the need for improved contextual tagging. This lack of contextual tagging significantly undermines the efficacy of CTI dissemination~\cite{Wagner2016MISP}.

Based on our analysis of platforms like the CIRCL MISPPriv platform, much of the sector-specific tagging is still performed manually by security analysts. This manual process is labor-intensive and prone to inconsistencies due to human error and varying levels of expertise among analysts. Automated tagging, using machine learning and natural language processing (NLP) techniques, can enhance the accuracy and consistency of tagging processes.

The effectiveness of CTI dissemination is critically dependent on the granularity and clarity of its sector categorization. While platforms like the CIRCL MISPPriv, our data source, employ a sector taxonomy with categories like 'government' and 'finance', the rationale for these categories is not always rigorously defined, leading to significant overlap and ambiguity. For example, a threat targeting a major bank could plausibly be tagged as 'finance', 'ICT', or even 'national security', depending on the analyst's interpretation. This inconsistency in manual tagging, stemming from an ill-defined taxonomy, is a major challenge that undermines the reliability of CTI dissemination. Our data-driven approach aims to mitigate this problem by learning the de facto relationships between threat events and sectors, thereby creating a more consistent and empirically-grounded tagging model.

\subsection{Problem Formulation}

The core challenge we address is the \textbf{automated sector-aware dissemination of Cyber Threat Intelligence (CTI)}. In practice, threat intelligence events are rarely tagged with the industry sectors (e.g., energy, finance, healthcare) that they impact. This lack of sector annotation significantly limits the operational value of CTI: organizations are left to manually filter vast amounts of generic intelligence to find what is relevant to them. For critical information infrastructure operators in particular, \textit{timely sector mapping is essential} to enable rapid defensive action against sector-targeted attacks.  

This problem is inherently a \textbf{multilabel classification task}. Unlike single-label classification, where each instance belongs to exactly one class, CTI events can simultaneously pertain to multiple sectors. For instance, a phishing campaign may exploit both healthcare and financial institutions, or an APT malware family may be observed targeting government and energy sectors at once. Off-the-shelf datasets are insufficient because they either lack sector labels altogether or provide incomplete and inconsistent tagging. Hence, a \textbf{specially crafted dataset} is required---one that accurately reflects sector relevance and is structured for machine learning.  

Formally, let a CTI event be represented as a textual description \( x \in \mathcal{X} \). The goal is to learn a function  

\begin{equation}
    f: \mathcal{X} \rightarrow \{0,1\}^K
\end{equation}

where \(K\) is the total number of industry sectors, and each output vector \( y = f(x) \) consists of binary entries \( y_k \in \{0,1\} \), with \( y_k = 1 \) indicating that the event \(x\) is relevant to sector \(k\). The multilabel nature of the task means that \(\sum_{k=1}^K y_k \geq 1\). 

Our formulation therefore centers on constructing a \textbf{sector-labeled CTI dataset} \(\{(x_i, y_i)\}_{i=1}^N\) from real-world Threat Intelligence Platforms (TIPs), where each \(x_i\) is a CTI feed entry, \(y_i\) is a sector-label vector, and \(N\) is the number of CTI events in the dataset. This dataset enables systematic evaluation of machine learning models for automated tagging, and provides the foundation for selecting the most effective approach for operational deployment. 

\section{Data Collection and Construction}\label{sec:data collection}
To automate the sector-specific tagging of CTI with machine learning, we must first address a fundamental challenge: the absence of a publicly available, labeled dataset suitable for this task. Existing CTI feeds are typically raw and unstructured, lacking the consistent, sector-specific labels required for supervised model training.

Therefore, the first problem we solve is one of data construction. In this section, we detail our methodology for creating a novel, sector-labeled CTI dataset. We outline the requirements for reliable CTI collection, describe our use of the MISPPriv\footnote{https://misppriv.circl.lu} platform, and explain the process of exporting, parsing, and processing raw feeds into a machine-learning-ready dataset of 872 tagged events. This section, therefore, outlines the step-by-step methodology for transforming raw CTI feeds into a structured, machine-learning-ready dataset.

\subsection{General Requirements for CTI Data Collection}
The threat intelligence data is collected from a diverse range of CTI feeds. 
There could be dozens or more feed sources available in the CTI communities. Therefore, we define several key requirements for collecting CTI data effectively:
\begin{enumerate}
    \item Efficiency. The data collection should be a streamlined process so that it could handle large volumes of data without significant delays.
    \item Source Reliability. The sources of CTI data must be credible and trustworthy to ensure the intelligence gathered is accurate and useful. 
    \item Data Diversity. A broad range of CTI sources should be used to capture a comprehensive view of the threat landscape. This includes different types of threats, indicators, and threat actor information.
\end{enumerate}   

\subsection{Selection of TIP}
To meet the previous requirements, we selected the Malware Information Sharing Platform for the Private Sector. Developed and maintained by a group of developers including Computer Incident Response Center Luxembourg (CIRCL), MISPPriv is designed to facilitate the collection and sharing of threat intelligence. 

MISPPriv satisfies the general requirements for CTI data collection:
\begin{itemize}
    \item Efficiency: MISPPriv allows for the automated collection of threat intelligence feeds, ensuring timely updates and minimal manual intervention.
    \item MISPPriv aggregates data from reputable sources, including national and international private organizations and CERTs (Computer Emergency Response Teams).
    \item The platform collects a wide range of threat intelligence, including Indicators of Compromise (IoCs), threat actor information, and other relevant data. Also, the platform collect CTI data from various sector. This feature could later support the CTI tagging model training process.
\end{itemize}

\begin{figure}[t]
    \centering
    \includegraphics[width=1.1\columnwidth]{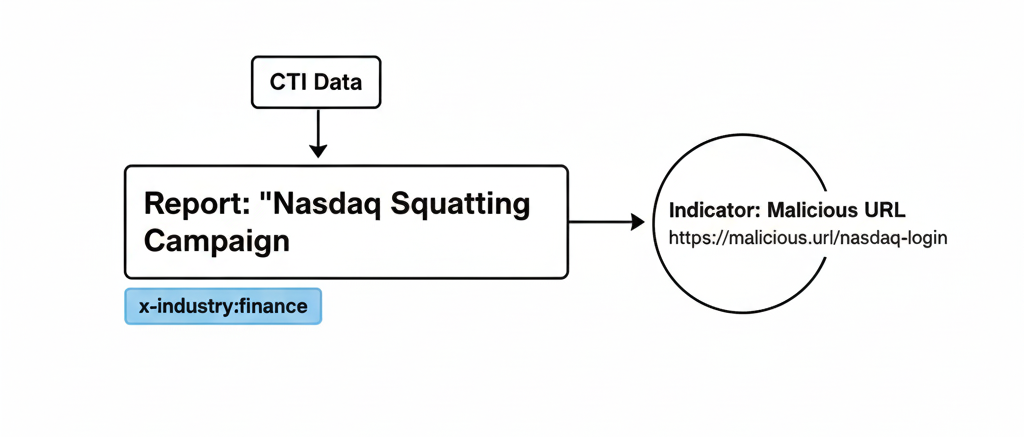}
    \caption{A high-level representation of a STIX 2.1 bundle. A \texttt{report} object is linked to an \texttt{indicator} object and enriched with a custom sector label (\texttt{x-industry:finance}) to provide context.}
    \label{fig:stix_example}
\end{figure}

This platform collects threat intelligence, Indicators of Compromise (IoCs), and threat actor information of any kind from CTI feeds using the STIX 2.1 format.
The STIX 2.1 format represents the IoC using a specific format called STIX Domain Objects (SDOs). SDOs have many categories that typically describe the tactics, techniques, and procedures (TTPs) of cyber incidents, such as Attack Pattern, Malware, Tools, and Vulnerabilities. There are at least 12 default SDOs categories that could be used to describe IoC. It is a common practice for different TIPs to develop their own custom SDOs for specific functions within their respective TIP frameworks. For example, the Sector SDO is not available by default. On the other hand, the Sector SDO is important for providing context to the cyber incident, making the CTI dissemination more accurate. Therefore, the MISP created \textit{misp-galaxy:sector} SDO to address this issue. This custom SDO ensures that the CTI data is tagged with relevant sector information, facilitating more effective dissemination. One such example of modified SDOs alongside the default one could be seen in~\Cref{fig:stix_example}.

\subsection{Data Collection and Processing using MISPPriv}

The Data Collection and Processing follows these specific steps:

\begin{enumerate}
    \item Exporting CTI Feeds. The data collection process begins with exporting the CTI feeds that are already sector-labeled from MISPPriv using the STIX 2.1 format.
    \item Parsing Data. The exported data is parsed using a custom script to extract attributes and specific SDOs relevant to the sector categorization process. These SDOs include Reports, Identity, Indicator, Tools, Threat Actor, Malware, Attack-Pattern, and Intrusion Set.
    \item Combining SDOs. In the parsing stage, all the relevant SDOs are combined into one property called IoC.
    \item Analyzing and Filtering. From approximately 300,000 feeds collected within the MISPPriv TIP as of December 2024, we constructed a labeled dataset of 872 diverse feeds, each labelled based on its specific critical sector. These sector-labeled feeds are essential for an accurate multilabel classification task.
\end{enumerate}

After parsing, the data is processed to extract relevant attributes and combine various SDOs into a unified IoC property. This process ensures that the CTI data is structured and ready for further analysis and dissemination.
\Cref{tab:1} shows the sample data of one CTI feed/event. By following these steps, the MISPPriv platform effectively collects and structures CTI data, meeting the general requirements for efficient, reliable, and diverse threat intelligence collection.

\begin{table*}[htbp]
    \centering
    \caption{Data Collection Output}
    \label{tab:1}
    \small
    \begin{tabular}{>{\centering\arraybackslash}p{2.8cm} 
    >{\centering\arraybackslash}p{1.8cm} 
    >{\centering\arraybackslash}p{2.3cm} 
    p{5.8cm} 
    >{\centering\arraybackslash}p{1.8cm}}
    \toprule
    \textbf{Report Name} & \textbf{Publisher} & \textbf{Published Date} & \textbf{IoC} & \textbf{Sector} \\ 
    \midrule
    OSINT: Chinese APT10 & dcso.de & 2021-10-04 & CnC; Redleaves benign DLL sideloading host; Compromised code-signing certificate, issued to Hacking Team & Government \\
    \bottomrule
    \end{tabular}
\end{table*}

\section{\DisCTI Design}

This section introduces the architecture of DisCTI, our proposed framework for automated sector-specific CTI dissemination. We describe the problem formulation, sector categorization, and the design of classification models, including binary classifier collections and a BERT-based multi-label classifier. What follows is a detailed description of our methodological approach, from data preprocessing steps to the specific architectures of the machine learning models used to automate the tagging process.

\subsection{Overview}

The purpose of this \DisCTI is to disseminate a targeted CTI for specific critical information infrastructure sectors. For example, the United States of America has defined as many as 16 critical infrastructure sectors that are further described in Presidential Policy Directive 21 (PPD-21). Japan, on the other hand, defined 15 sectors as CII (Critical Information Infrastructure), including ICT, finance, airport, and medical services~\cite{CybersecurityStrategicHQ2022CIP}. 

The Indonesian Government defines the CII sectors in Presidential Regulation No 82:2022 titled Vital Information Infrastructure Protection Act. There are 8 sectors including government, defense, transportation, finance, health, energy, ICT, and agriculture~\cite{PIIVreg2022}. In this work, CTI dissemination is categorized based on the 8 sectors mentioned earlier plus 3 additional sectors: industry, education, and media. Industry and education are added because a significant volume of CTI information relates to those sectors. Media is added on operational grounds: within BSSN, the directorate responsible for the ICT sector also holds the mandate for the media and transportation sectors, so media-targeted intelligence must be routed through the same dissemination pathway. This yields the 11 target sectors used throughout the remainder of this paper.

\begin{figure}[t]
    \centering
    \includegraphics[width=0.9\columnwidth]{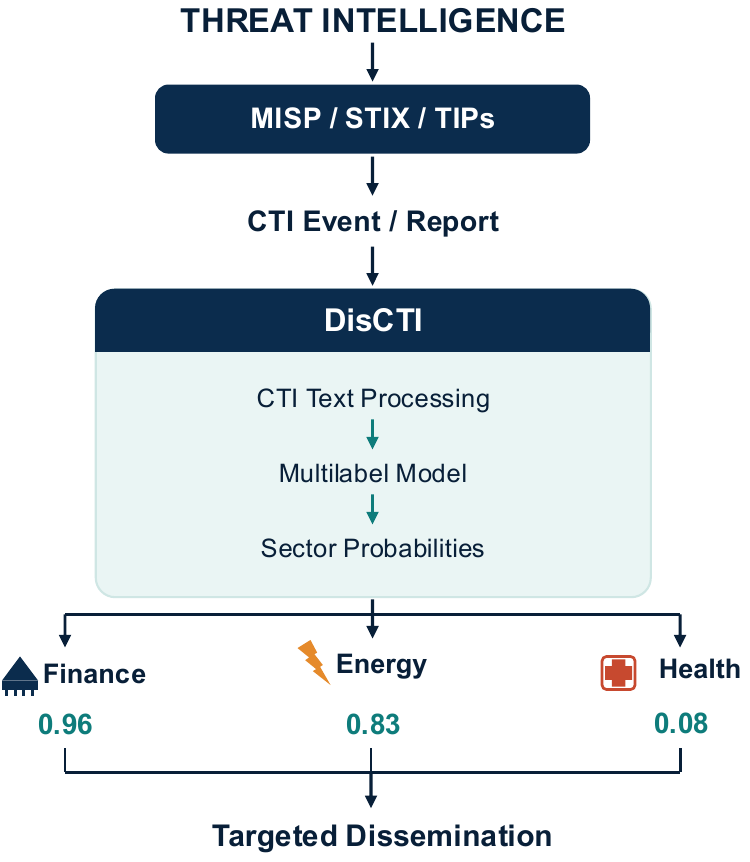}
    \caption{A high-level overview of DisCTI System Design.}
    \label{fig:threat_tagging}
\end{figure}

Meanwhile, the model output is in the form of categorized CTI based on CII sectors as depicted in Fig~\ref{fig:threat_tagging}. The dataset, presented in a specific format at~\Cref{tab:1}, is used to train the model. Overall, the \DisCTI needs to be resolved as a multi-label classification or tagging problem. For each CTI sample data, the sector to which the CTI belongs may be more than one sector. 

To address this tagging challenge, we propose solutions that formulate the problem as a multilabel classification task. We use two approaches from this method. The first solution employs a technique called the Problem Transformation Method, which utilises a collection of binary classifiers, each corresponding to a sector. The second approach is to use a single model that performs multi-label classification immediately. Regardless of the approach, our pipeline involves the steps below:

\begin{enumerate}
    \item Data Preprocessing: This involves normalizing whitespace and splitting input text into sentences, inserting special tokens for BERT processing~\cite{devlin2019bert}.
    \item Tokenization: The data preparation process within  Natural Language Processing (NLP) which breaks down the input text into smaller units, called tokens. In the Binary Classifier, the tokens are produced by splitting the text into words or sub words. For BERT, the text will be converted into input IDs, attention masks, and token type IDs~\cite{wolf2020transformers}.
    \item Data Formatting for PyTorch: Formatting tokenized data as PyTorch tensors for model input during training and evaluation \cite{Paszke2019PyTorch}.
\end{enumerate}

These steps are crucial for preparing the raw CTI data for processing by the models we used, ensuring that the model can effectively learn from and classify the data.

\subsection{Binary Classifier Collection}

The collection of binary classifiers can be organized parallelly/independently and sequentially/dependently.

\subsubsection{\noindent\textbf{Parallel Binary Classifiers.}}
This organization treats each label as an independent binary classification problem, training a separate classifier for each label \cite{Zhang2014}. While simpler to implement, the Parallel Binary Classifiers model does not consider label dependencies, which can be a limitation in some contexts.

\subsubsection{\noindent\textbf{Sequential Binary Classifier.}}
This model uses a chain of binary classifiers, each responsible for predicting a single label, with the predictions of previous classifiers being used as additional features for subsequent classifiers \cite{Read2011}. The Sequential Binary Classifiers model can capture label dependencies, thereby improving multilabel classification performance.

\subsection{BERT based Multi-Label Classifier}
Our other approach employs BERT (Bidirectional Encoder Representations from Transformers), a pre-trained transformer-based model known for its excellence in natural language processing tasks~\cite{devlin2019bert}. A transformer is a deep learning architecture developed by Google and based on the multi-head attention mechanism.  By using BERT, we aim to automate the tagging process and improve the relevance and utility of CTI feeds.

However, it is important to note that while BERT offers superior performance, it also might incur computational cost, which can serve as a baseline for future work. In our experiments, training the model for 25 epochs took approximately 125 minutes (5 minutes per epoch) on a T4 GPU with 15GB of VRAM. This highlights that, while effective, the memory usage and running time of BERT-based models are considerably higher than those of simpler models due to their large number of parameters and multi-head attention mechanism. Therefore, the efficiency of this approach must be evaluated against available hardware resources, particularly when scaling to larger datasets or real-time applications.

\subsection{Implementation: Data Preprocessing}
The representation of the raw dataset like the one shown in~\Cref{tab:1} needs to be further processed to be suitable for NLP model training. Data preprocessing consists of two stages: label encoding, and IoC data cleaning. 

\subsubsection{\noindent{\bf Label Encoding.}} The label encoding stage is a fundamental step in preparing the data for a multilabel classification problem, especially when leveraging the BERT model. The encoding will be done based on the result of unique label identification. After we define the total unique label set based on critical sectors, the encoding process converts textual or categorical labels into numerical values that can be processed by the models. This transformation is crucial because the models require numerical inputs for both features and labels. There are at least two main ideas in this stage: 
\begin{itemize}
    \item Identification of unique labels. There are multiple unique labels present in the dataset. These labels correspond to different CII sectors that need to be tagged. 
    \item Binary Encoding. Given the multilabel nature of this problem, each data point can be associated with multiple labels. Therefore, we use a binary encoding method in which each unique label is represented as a binary vector. In this vector, each position corresponds to a specific label; the value is 1 if the label is present and 0 if it is not. As an example, if in a sample assumption there are only three labels like 'Transportation', 'Healthcare', and 'Finance', then a data point relevant to 'Transportation' sector would be encoded as [1, 0, 0].  
\end{itemize}

After the label encoding stage, we transform the structure of~\Cref{tab:1} by concatenating the 'Report Name', 'Publisher', and the 'IoC' columns into one 'event info' column. Meanwhile, since the 'Published Date' column is not needed to classify CTIs by related sectors, it is dismissed. 

\subsubsection{\noindent{\bf IoC Data Cleaning.}} The next stage is cleaning the IoC data from unnecessary markings and punctuation. The purpose is to ensure that the data fed into the models is high-quality and free of noise. This step is needed because raw CTI data typically contains various forms of noise and inconsistencies, such as special characters, redundant whitespace, URLs, and other non-informative elements. 

The text cleaning process involves several steps: 
\begin{itemize}
    \item Normalization of whitespace.  
    \item Removal of Special Characters and Non-Word characters.
    \item Lowercase. This stage will convert all letters to lowercase. This is crucial because the important part of CTI understanding is not related to uppercase/lowercase.
    \item Lemmatization. We define a function that applies the lemmatizer to each word in a sentence. This step standardizes the input data, making it more uniform and reducing the vocabulary size.
\end{itemize}

After cleaning, the dataset is split into a training and a testing dataset at a 75:25 ratio. 
This means that from 872 Threat Intelligence records, we get roughly 654 CTI events for training and 218 for validation. 

\subsection{Training: Parallel Binary Classifiers and Sequential Binary Classifiers}
Parallel Binary Classifiers and Sequential Binary Classifiers are two approaches to multilabel classification. In Parallel Binary Classifiers, separate binary classifiers are trained independently for each label, while in Sequential Binary Classifiers, classifiers are linked in a chain, with each classifier considering the predictions of the previous classifiers as additional features. Both approaches require the textual data to be vectorized in numerical form for effective training and prediction.

The TF-IDF vectorizer transforms the text data into numerical features by considering both the frequency of terms in a document and their inverse frequency across all documents \cite{Manning2008}. We run two steps in this regard as shown in~\Cref{alg:tfidf}:
\begin{itemize}
    \item Initialization of TF-IDF Vectorizer. This initialization of TF-IDF vectorizer is obtained from the scikit-learn library.
    \item Fitting and Transforming the Data. The TF-IDF vectorizer is then fitted to the cleaned and lemmatized text data. The text data is transformed into TF-IDF vectors, resulting in a numerical representation of the text. The fitting stage is done based on the training dataset, and the transform stage is done for both training and validation dataset. 
\end{itemize}

\begin{algorithm}[!t]
\caption{TF-IDF Vectorization}
\label{alg:tfidf}
\begin{algorithmic}[1]
\Inputs{\texttt{train.english\_event} (training text data), \texttt{train} (dataframe containing labels)}
\Outputs{\texttt{x\_train} (vectorized text), \texttt{y\_train} (binary label vectors)}
\Initialize{
  \texttt{strip\_accents} = 'unicode', 
  \texttt{analyzer} = 'word', 
  \texttt{ngram\_range} = (1, 1), 
  \texttt{norm} = 'l2'
}
\State Initialize TF-IDF vectorizer with above parameters
\State Fit vectorizer on training text:
\Statex \hspace{1em} \texttt{vectorizer.fit(train.english\_event)}
\State Transform text to TF-IDF vectors:
\Statex \hspace{1em} \texttt{x\_train = vectorizer.transform(train.}
        \hspace{1em} \texttt{english\_event)}
\State Extract label matrix from dataframe:
\Statex \hspace{1em} \texttt{y\_train = train[[\textquotesingle Agriculture\textquotesingle, \textquotesingle Defense\textquotesingle, ..., \textquotesingle Transport\textquotesingle]]}
\end{algorithmic}
\end{algorithm}

\subsubsection{\noindent{\bf{Parallel Binary Classifiers Training Stage}}}
Following the vectorization of the text data using the TF-IDF vectorizer, the next step is to train the models. In the Parallel Binary Classifiers approach, each label is treated as an independent binary classification problem. Separate classifiers are trained for each label, which makes this approach straightforward and scalable. 

\begin{algorithm}[H]
\caption{Parallel Binary Classifiers Training}
\label{alg:binary-relevance}
\begin{algorithmic}[1]
\Inputs{ Training data (\texttt{x\_train}, \texttt{y\_train})}
\Outputs{ Prediction (\texttt{prediction})}
\raggedright
\State Import \texttt{BinaryRelevance} from \texttt{skmultilearn.problem\_transform}
\State Import \texttt{GaussianNB} from \texttt{sklearn.naive\_bayes}
\State Initialize \texttt{classifier} as \texttt{BinaryRelevance(GaussianNB())}
\State Fit \texttt{classifier} on the training data: \texttt{classifier.fit(x\_train, y\_train)}
\State Predict on the test data: \texttt{prediction = classifier.predict(x\_test)}
\end{algorithmic}
\end{algorithm}

For each label, a Gaussian Naive Bayes classifier is used. Gaussian Naive Bayes is chosen for its simplicity and effectiveness in handling high-dimensional data. The TF-IDF vectors obtained during preprocessing are used as input features. The BinaryRelevance classifier from skmultilearn.problem is used to fit multiple Gaussian Naive Bayes classifiers, one for each label \cite{Szymański2017}. During the prediction phase, each classifier independently predicts the presence or absence of its corresponding label. We implement all of those stages as described in~\Cref{alg:binary-relevance}. 

\subsubsection{\noindent{\bf{Sequential Binary Classifiers Training Stage}}}
In the Sequential Binary Classifiers approach, classifiers are linked in a chain, where each classifier considers the predictions of the previous classifiers as additional features. This approach captures label dependencies, which can improve the overall performance.

A Random Forest Classifier is chosen for its robustness and ability to handle large feature sets. The ClassifierChain wrapper is used to link the classifiers. Same with the Parallel Binary Classifiers, the TF-IDF vectors are used as input features. The ClassifierChain is fitted with a Random Forest Classifier, and the chain order is determined by the sequence of labels. Finally, to make predictions, the classifiers in the chain make predictions sequentially, with each classifier taking into account the predictions of its predecessors like what is shown by~\Cref{alg:classifier-chain}. 

\begin{algorithm}[!t]
\caption{Sequential Binary Classifiers Training}
\label{alg:classifier-chain}
\begin{algorithmic}[1]
\Inputs{ Training data (\texttt{x\_train}, \texttt{y\_train})}
\Outputs{ Prediction (\texttt{prediction})}
\raggedright
\State Import \texttt{ClassifierChain} from \texttt{skmultilearn.problem\_transform}
\State Import \texttt{RandomForestClassifier} from \texttt{sklearn.ensemble}
\State Initialize \texttt{classifier} as \texttt{ClassifierChain(RandomForest Classifier (n\_estimators=150))}
\State Fit \texttt{classifier} on the training data: \texttt{classifier.fit(x\_train, y\_train)}
\State Predict on the test data: \texttt{prediction = classifier.predict(x\_test)}
\end{algorithmic}
\end{algorithm}

\subsection{Training: BERT Model}
When preparing data for a BERT model, the tokenization and embedding process are critical steps that convert raw text into numerical data suitable for model training. This section describes these processes in detail. 

\subsubsection{\noindent{\bf{Tokenization}}}
BERT uses a WordPiece tokenizer, which breaks down words into subword units. This approach is particularly effective in handling out-of-vocabulary words like what we have inside most of CTI and capturing nuanced meanings in CTI data.

\begin{algorithm}[!t]
\caption{BERT Tokenization and Embedding}
\label{alg:bert-tokenization}
\begin{algorithmic}[1]
\Inputs{ Text data (\texttt{dataframe['english\_event']})}

\Outputs{ Tokenized inputs}

\Initialize{ \texttt{tokenizer} using \texttt{BertTokenizer.from\_pretrained ('bert-base-uncased')}}

\raggedright
\State Import \texttt{BertTokenizer} from \texttt{transformers}
\State Tokenize the input text:
\Statex \hspace{1em} \texttt{inputs = tokenizer.encode\_plus( dataframe['english\_event'],}
\Statex \hspace{4em} \texttt{add\_special\_tokens=True,}
\Statex \hspace{4em} \texttt{max\_length=self.max\_len,}
\Statex \hspace{4em} \texttt{padding='max\_length',}
\Statex \hspace{4em} \texttt{return\_token\_type\_ids=True,}
\Statex \hspace{4em} \texttt{truncation=True)}

\State Convert inputs to tensors:
\Statex \hspace{1em} \texttt{'ids': torch.tensor(inputs ['input\_ids'], dtype=torch.long)}
\Statex \hspace{1em} \texttt{'mask': torch.tensor(inputs ['attention\_mask'], dtype=torch.long)}
\Statex \hspace{1em} \texttt{'token\_type\_ids':torch.tensor(
inputs ["token\_type\_ids"], dtype=torch.long)}
\Statex \hspace{1em} \texttt{'targets': torch.tensor(
self.targets[index], dtype=torch.float)}
\end{algorithmic}
\end{algorithm}

\begin{itemize}
    \item Initialization.
    The BertTokenizer from the transformers library by Hugging Face is initialized using a pre-trained BERT model. This tokenizer is pre-configured with the vocabulary and rules specific to BERT.
    \item Tokenization. The tokenizer is applied to the text data, converting each document into a sequence of tokens. Special tokens [CLS] and [SEP] are added at the beginning and end of each sequence to denote the start and end of the input, respectively. The [PAD] special tokens are also added to fulfill the length requirement of the BERT-base model which is 512 tokens. \Cref{alg:bert-tokenization} explains the implementation phase of this Tokenization.     
\end{itemize}

\subsubsection{\noindent{\bf{Embedding}}}
The embedding process involves converting the tokenized text into dense vectors of fixed size that capture the semantic information of the tokens. BERT generates these embeddings using its pre-trained transformer architecture.
\begin{itemize}
    \item Creating Attention Masks. Attention masks are created to differentiate between actual tokens and padding tokens. 
    \item Tensor Conversion. The input IDs and attention masks are converted into PyTorch tensors, which are the required input format for the BERT model in the transformers library.
    \item Embedding Generation. During training, BERT processes the previous tensors through its layers, generating contextualized embeddings for each token. This stage will be described later in the training stage. 
\end{itemize}

The tokenization and embedding process is integrated into the BERT model training pipeline. The resulting embeddings are fed into the BERT model, fine-tuned for the specific task of CTI dissemination. \Cref{alg:bert-tokenization} shows the overall process of tokenization and embedding. 

\subsubsection{\noindent{\bf{BERT Training Stage}}}
\begin{figure}[h]
    \centering
    \includegraphics[width=1\columnwidth]{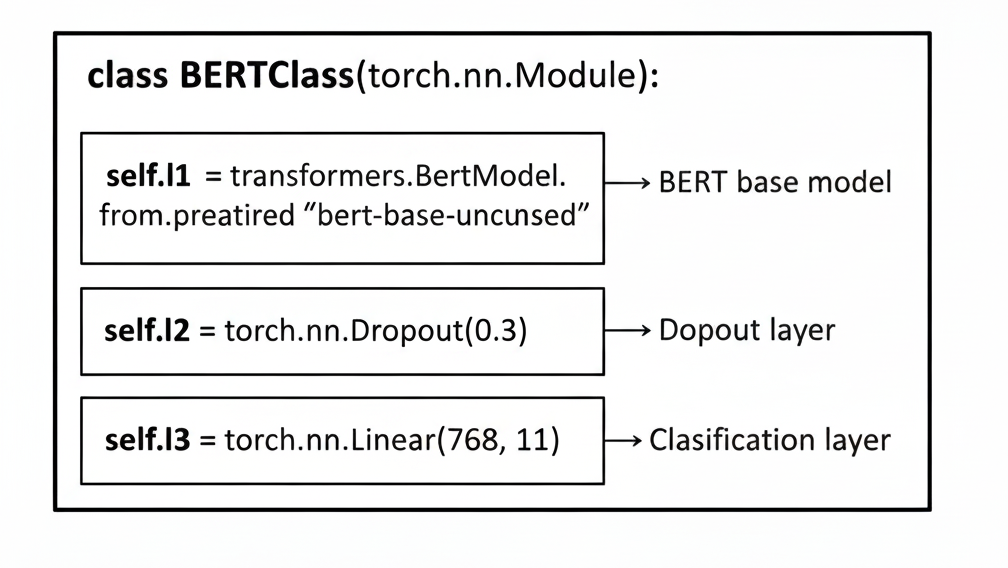}
    \caption{The architecture of our BERT-based multi-label classifier.}
    \label{fig:bert_architecture}
\end{figure}

This section outlines the important steps involved in training the BERT model, including model configuration, training parameter configuration, and the actual training process. Model configuration are shown in~\Cref{fig:bert_architecture}. 

\begin{itemize}
    \item Model Initialization. 
    The BERT model is initialized with a pre-trained version from the transformers library. The model's configuration is adjusted to handle multilabel classification by setting the number of output labels to match the number of classes in the dataset.
    \item Optimizer and Learning Rate. The AdamW optimizer is chosen for its ability to handle large-scale datasets and its suitability for transformer models \cite{Kingma2014Adam}. A linear learning rate scheduler is employed to gradually adjust the learning rate during training, improving convergence and preventing overfitting.
    \item Model Training Preparation. The dataset is divided into training and validation sets. DataLoaders are created for efficient batch processing during training and evaluation \cite{Muller2019}. 
    \item Training loop. The training loop iterates over the specified number of epochs which is 25 epochs. During each epoch, the model processes batches of data which is 6 in size, computes the loss, and updates the model parameters. The loss function used is binary cross-entropy (BCE), suitable for multilabel classification.
    \item Validation. After each epoch, the model's performance is evaluated on the validation set to monitor its progress and adjust hyperparameters if necessary. The validation loop is similar to the training loop but excludes the backpropagation step.
\end{itemize}

\subsection{Evaluation Metrics}

The performance of the models is evaluated using standard multilabel classification metrics. These metrics provide a comprehensive understanding of how well the models perform across multiple labels.

\subsubsection{Hamming Loss}
Hamming loss measures the fraction of labels that are incorrectly predicted \cite{Schapire2000}. It is defined as: 
\[
\text{Hamming Loss} = \frac{1}{nL} \sum_{i=1}^{n} \sum_{j=1}^{L} \text{I}(\hat{y}_{ij} \neq y_{ij})
\]
where \( n \) is the number of samples, \( L \) is the number of labels, \( \hat{y}_{ij} \) is the predicted label, and \( y_{ij} \) is the true label. The \(I\) is an Indicator function that is used to evaluate whether a specific condition is true or false.

Throughout this paper, we use \textit{label-wise accuracy} to denote \(1 - \text{Hamming Loss}\), i.e.\ the fraction of individual sector-label assignments predicted correctly across all samples and labels. This is distinct from \textit{exact-match accuracy} (also called subset accuracy), which counts a sample as correct only when all \(L\) of its sector labels are predicted correctly simultaneously, and which is therefore substantially lower for any multilabel task. We report the label-wise variant because it reflects the operational cost structure of dissemination, where each sector-routing decision is acted upon independently; we state the metric explicitly at each use to avoid ambiguity.

\subsubsection{Precision}
Precision measures the proportion of true positive predictions among all positive predictions made by the model \cite{Sokolova2009}. It is defined as:
\[
\text{Precision} = \frac{\text{True Positives}}{\text{True Positives} + \text{False Positives}}
\]

\subsubsection{Recall}
Recall measures the proportion of true positive predictions among all actual positives in the dataset \cite{Sokolova2009}. It is defined as:
\[
\text{Recall} = \frac{\text{True Positives}}{\text{True Positives} + \text{False Negatives}}
\]

\subsubsection{F1-Score}
The F1-score is the harmonic mean of precision and recall, providing a single metric that balances both precision and recall \cite{Sokolova2009}. It is defined as:
\[
\text{F1-Score} = 2 \times \frac{\text{Precision} \times \text{Recall}}{\text{Precision} + \text{Recall}}
\]

\section{Experiments}

\subsection{Overview and Research Questions} 
Experiments aim at answering the following research questions (RQs):

\begin{itemize}
    \item \textbf{RQ 1.} How can sector-specific tagging of CTI data be automated to improve the relevance of disseminated information?
    
    We explored rule-based and machine learning methods, including a BERT-based multi-label classifier, to automate sector-specific tagging of Cyber Threat Intelligence (CTI) data, improving information relevance and timeliness. The BERT model demonstrated decent accuracy in tagging sector-specific threats, with performance evaluated using metrics such as precision, recall, and F1 score. 
    
    \item \textbf{RQ 2.} Which machine learning model provides the best performance? 
    
    To determine the best machine learning model for automating sector-specific tagging of CTI data, we first explored binary classifiers. While they performed adequately in single-label classification, they struggled with the multi-sector nature of CTI data. Multi-label binary classifiers improved results slightly but were still limited. In contrast, the BERT-based multi-label classifier provided superior performance, effectively handling multiple sector tags simultaneously and achieving better precision, recall, and F1 scores. Therefore, while binary classifiers can address simplified cases, multi-label models like BERT are more effective for this complex task.
\end{itemize}

\subsection{Experimental 
Result: Parallel Binary Classifiers with Gaussian Naive Bayes}
After training the models, the next critical step is to evaluate their performance. 
The Parallel Binary Classifiers model treats each label independently, leveraging Gaussian Naive Bayes classifiers. The results indicate that while this approach is simple and scalable, it does not capture label dependencies, which can limit its performance.

\begin{table}[htbp]
    \centering
    \caption{Evaluation metrics for parallel binary classifiers model using Gaussian Naive Bayes}
    \label{tab:2}
    \footnotesize
    \setlength{\tabcolsep}{3pt}
    \begin{tabularx}{\linewidth}{lCCCC}
        \toprule
        \textbf{Label} & \textbf{Precision} & \textbf{Recall} & \textbf{F1-Score} & \textbf{Support} \\
        \midrule
        Agriculture     & 0.889 & 1.000 & 0.941 & 8   \\
        Defense         & 0.523 & 0.719 & 0.605 & 32  \\
        Education       & 0.733 & 0.805 & 0.767 & 41  \\
        Energy          & 0.730 & 0.885 & 0.800 & 61  \\
        Finance         & 0.667 & 0.774 & 0.716 & 62  \\
        Government      & 0.793 & 0.823 & 0.807 & 79  \\
        Health          & 0.768 & 0.843 & 0.804 & 51  \\
        ICT             & 0.811 & 0.827 & 0.819 & 63  \\
        Industrial      & 0.718 & 0.884 & 0.792 & 69  \\
        Media           & 0.688 & 1.000 & 0.815 & 11  \\
        Transport       & 0.875 & 0.860 & 0.867 & 57  \\
        \midrule
        \textbf{Micro Avg}     & \textbf{0.740} & \textbf{0.837} & \textbf{0.786} & \textbf{534} \\
        \textbf{Macro Avg}     & \textbf{0.745} & \textbf{0.856} & \textbf{0.794} & \textbf{534} \\
        \textbf{Weighted Avg}  & \textbf{0.747} & \textbf{0.837} & \textbf{0.788} & \textbf{534} \\
        \textbf{Samples Avg}   & \textbf{0.640} & \textbf{0.712} & \textbf{0.647} & \textbf{534} \\
        \bottomrule
    \end{tabularx}
\end{table}

The results based on~\Cref{tab:2} indicate that the Parallel Binary Classifiers model using Gaussian Naive Bayes performs reasonably well across most categories. The precision, recall, and F1-scores for each sector provide insights into the model's effectiveness in identifying true positives and minimizing false positives and negatives.

\begin{enumerate}
    \item Precision: The model achieved high precision in the Agriculture (0.889) sector, indicating a lower false-positive rate. However, precision is lower in sectors like Defense (0.523), suggesting room for improvement in predicting true positive instances accurately.
    \item Recall: The recall scores indicate the model's ability to identify all relevant instances. Sectors like Agriculture (1.000) achieved perfect recall, while others like Defense (0.719) showed lower recall, indicating missed instances.
    \item F1-Score: The F1-scores provide a balanced measure of the model's accuracy. The Agriculture sector had the highest F1-score (0.941), reflecting strong precision and recall. Defense had the lowest F1-score (0.605), indicating the need for better balance between precision and recall.
    \item The Parallel Binary Classifiers model's overall accuracy is 0.9158, indicating that approximately 91.58\% of the labels were correctly predicted. The Hamming Loss of 0.0842 reflects the fraction of incorrect labels to the total number of labels, indicating a relatively low error rate.
\end{enumerate}

\subsection{Experimental Result: Sequential Binary Classifiers with Random Forest Classifier}
We present the results and analysis of our Sequential Binary Classifiers approach using the Random Forest Classifier. This model accounts for interdependencies among labels, potentially improving performance in multi-label classification tasks compared to simpler methods such as Parallel Binary Classifiers.

\Cref{tab:3} presents the detailed evaluation metrics for each sector, including micro, macro, weighted, and sample averages. These results provide insights into how well the model performs across different sectors and the overall performance when considering all instances and labels collectively.

\begin{table}[htbp]
    \centering
    \caption{Evaluation metrics for sequential binary classifiers models using Random Forest classifier}
    \label{tab:3}
    \footnotesize
    \setlength{\tabcolsep}{3pt}
    \begin{tabularx}{\linewidth}{lCCCC}
        \toprule
        \textbf{Label} & \textbf{Precision} & \textbf{Recall} & \textbf{F1-Score} & \textbf{Support} \\
        \midrule
        Agriculture     & 1.000 & 0.875 & 0.933 & 8   \\
        Defense         & 1.000 & 0.625 & 0.769 & 32  \\
        Education       & 0.939 & 0.756 & 0.838 & 41  \\
        Energy          & 0.850 & 0.836 & 0.843 & 61  \\
        Finance         & 0.894 & 0.677 & 0.771 & 62  \\
        Government      & 0.889 & 0.709 & 0.789 & 79  \\
        Health          & 0.933 & 0.824 & 0.875 & 51  \\
        ICT             & 0.953 & 0.788 & 0.863 & 63  \\
        Industrial      & 0.925 & 0.710 & 0.803 & 69  \\
        Media           & 1.000 & 0.727 & 0.842 & 11  \\
        Transport       & 0.964 & 0.930 & 0.946 & 57  \\
        \midrule
        \textbf{Micro Avg}     & \textbf{0.922} & \textbf{0.765} & \textbf{0.836} & \textbf{534} \\
        \textbf{Macro Avg}     & \textbf{0.941} & \textbf{0.769} & \textbf{0.843} & \textbf{534} \\
        \textbf{Weighted Avg}  & \textbf{0.923} & \textbf{0.765} & \textbf{0.834} & \textbf{534} \\
        \textbf{Samples Avg}   & \textbf{0.614} & \textbf{0.602} & \textbf{0.599} & \textbf{534} \\
        \bottomrule
    \end{tabularx}
\end{table}

The Sequential Binary Classifiers model, utilizing a Random Forest Classifier, showed robust performance across multiple metrics:

\begin{enumerate}
    \item Precision: The model achieved high precision in most categories, with perfect precision scores (1.000) in Agriculture and Defense, indicating a low rate of false positives in these sectors.
    \item Recall: The recall scores varied, with sectors like Energy (0.836) and Transport (0.930) performing well, while Defense had a lower recall (0.625). This variation suggests the model's varying effectiveness in identifying all relevant instances across different sectors.
    \item F1-Score: The F1-scores balanced the precision and recall, providing an overall measure of the model's accuracy. The Transport sector achieved the highest F1-score (0.946), reflecting balanced precision and recall. Conversely, the Defense sector had a lower F1-score (0.769), indicating a need for improvement in balancing precision and recall.
    \item The model's label-wise accuracy is 0.9324, i.e.\ a Hamming Loss of 0.0676, which is slightly better than the Parallel Binary Classifiers model.
\end{enumerate} 

Comparing the Sequential Binary Classifiers results with the Parallel Binary Classifiers approach reveals a consistent trade-off: the Sequential model attains markedly higher precision across nearly every sector, at the cost of lower recall. For instance, the Agriculture sector achieved a perfect precision score with the Sequential Binary Classifiers model, whereas the Parallel Binary Classifiers model's precision was slightly lower at 0.889. The same pattern holds for the Government sector, where precision improved from 0.793 (Parallel) to 0.889 (Sequential) while recall fell from 0.823 to 0.709.

This trade-off is visible at the aggregate level as well: macro-averaged precision rises from 0.745 to 0.941 while macro-averaged recall falls from 0.856 to 0.769. The Defense sector illustrates the cost most sharply, with recall dropping from 0.719 (Parallel) to 0.625 (Sequential), suggesting that the Parallel Binary Classifiers approach may remain preferable where recall is operationally more important than precision---as is often the case in threat dissemination, where a missed sector is typically costlier than a spurious one.

Overall, the Sequential Binary Classifiers model, with its high precision and balanced recall, offers a more comprehensive and reliable method for multi-label classification in the context of CTI dissemination.

\begin{figure}[ht]
  \centering
  \includegraphics[width=0.5\textwidth]{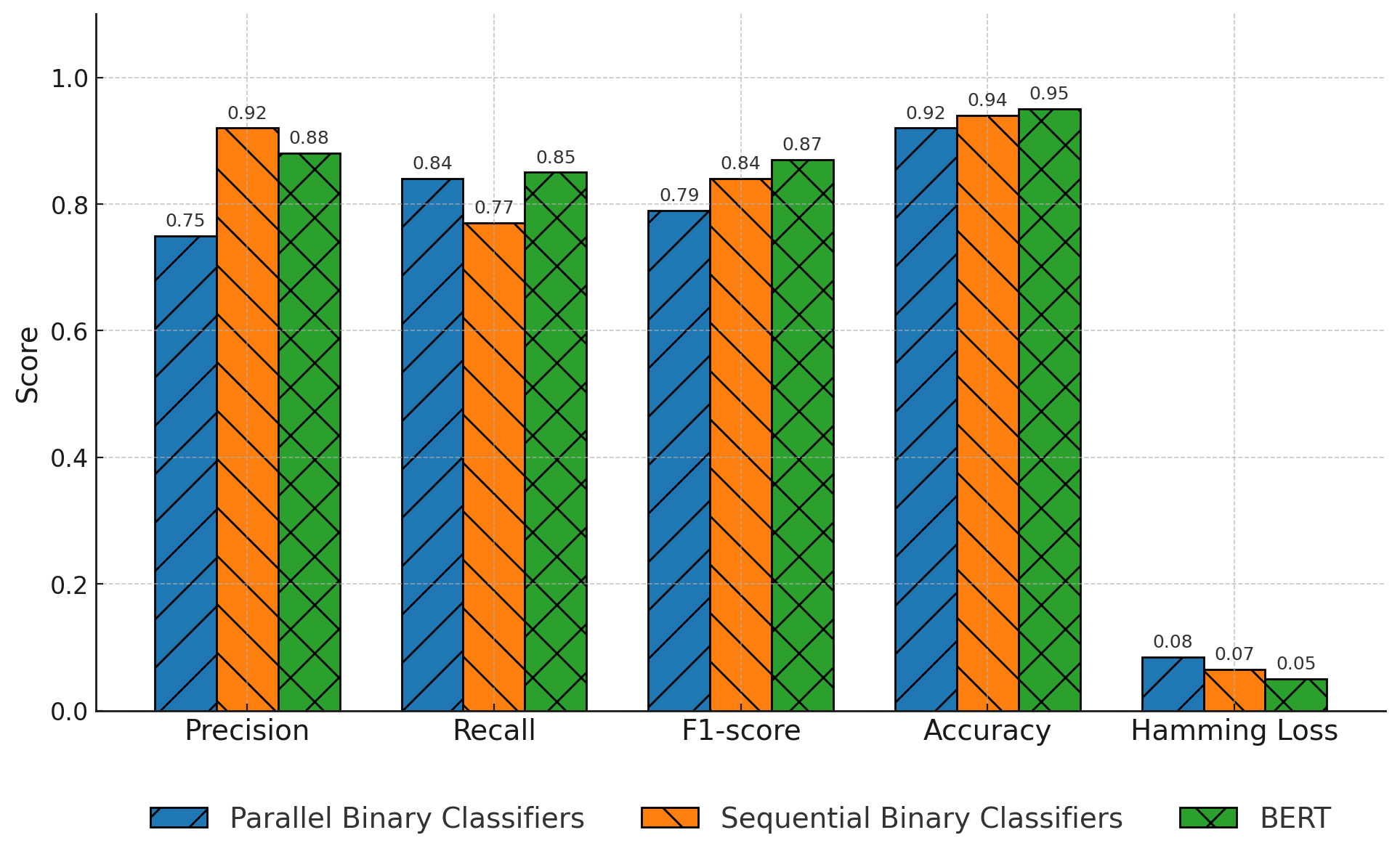}
  \caption{Comparative analysis of evaluation metrics for Parallel Binary Classifiers, Sequential Binary Classifiers, and BERT models. Metrics include Precision, Recall, F1-score, Accuracy, and Hamming Loss. The BERT model consistently outperforms the other approaches across most metrics.}
  \label{fig:comparison}
\end{figure}

\subsection{Experimental Result: BERT Model}
The BERT model, leveraging a transformer-based architecture, is evaluated on the same multi-label classification task.~\Cref{tab:4} presents the detailed evaluation metrics for each sector, providing a comprehensive overview of the model's performance across different labels.

\begin{table}[htbp]
    \centering
    \caption{Evaluation metrics for BERT model. Averages are computed over the
    10 sectors listed; the media sector is not included in this evaluation (see
    the note at the end of this subsection).}
    \label{tab:4}
    \footnotesize
    \setlength{\tabcolsep}{3pt}
    \begin{tabularx}{\linewidth}{lCCCC}
        \toprule
        \textbf{Label} & \textbf{Precision} & \textbf{Recall} & \textbf{F1-Score} & \textbf{Support} \\
        \midrule
        Agriculture  & 1.000 & 1.000 & 1.000 & 10 \\
        Defense      & 1.000 & 0.680 & 0.810 & 37 \\
        Education    & 0.830 & 0.810 & 0.820 & 42 \\
        Energy       & 0.870 & 0.930 & 0.900 & 59 \\
        Finance      & 0.870 & 0.830 & 0.850 & 65 \\
        Government   & 0.910 & 0.840 & 0.880 & 88 \\
        Health       & 1.000 & 0.940 & 0.970 & 53 \\
        ICT          & 0.950 & 0.850 & 0.890 & 65 \\
        Industrial   & 0.850 & 0.770 & 0.810 & 71 \\
        Transport    & 0.960 & 0.950 & 0.950 & 56 \\
        \midrule
        \textbf{Micro Avg}     & \textbf{0.910} & \textbf{0.850} & \textbf{0.880} & \textbf{546} \\
        \textbf{Macro Avg}     & \textbf{0.920} & \textbf{0.860} & \textbf{0.890} & \textbf{546} \\
        \textbf{Weighted Avg}  & \textbf{0.910} & \textbf{0.850} & \textbf{0.880} & \textbf{546} \\
        \textbf{Samples Avg}   & \textbf{0.750} & \textbf{0.740} & \textbf{0.740} & \textbf{546} \\
        \bottomrule
    \end{tabularx}
\end{table}

The BERT model exhibits strong performance across several metrics, demonstrating the potential advantages of using transformer-based architectures for multi-label classification tasks:

\begin{enumerate}
    \item Precision: The model achieved perfect precision (1.000) in Agriculture, Defense, and Health sectors, indicating no false positives in these categories. High-precision scores were also observed in ICT (0.950) and Transport (0.960), suggesting that the BERT model is effective at correctly identifying true positives. 
    \item Recall: Recall scores varied, with sectors like Agriculture (1.000) and Transport (0.950) performing well, while Defense had a lower recall (0.680), indicating some missed instances.
    \item F1-Score: The F1-scores provide a balanced measure of the model's accuracy. The Agriculture sector achieved a perfect F1 Score (1.000), indicating strong precision and recall. Conversely, Defense and Industry sectors had a lower F1-score (0.810), indicating a need for better balance between precision and recall.
    \item The BERT model's overall Hamming Loss of 0.055 reflects the fraction of incorrect labels to the total number of labels, indicating a relatively low error rate. Equivalently, 94.5\% of individual sector-label assignments are correct. We report this label-wise figure rather than exact-match accuracy, which is a considerably stricter criterion in multilabel settings because it requires every one of a sample's sector labels to be predicted correctly simultaneously.
\end{enumerate}

\subheading{Scope of the BERT evaluation} The BERT results in~\Cref{tab:4} cover 10 of the 11 sectors defined in Section~\ref{sec:related work}; the media sector, which carries the smallest label support in the dataset (11 instances, see~\Cref{tab:2}), is not included. Consequently the averages in~\Cref{tab:4} are not strictly comparable with those in~\Cref{tab:2} and~\Cref{tab:3}, which are computed over all 11 sectors. We flag this explicitly rather than silently omitting the sector, and note that a like-for-like re-evaluation of all three models over an identical 11-sector split is in progress as part of ongoing work; the qualitative ranking of the three approaches is not expected to change, since media contributes roughly 2\% of label instances.

\subsection{Comparative Analysis}
In this section, we present the results of a comparative analysis between the three models. We compare the performance of three models for targeted CTI dissemination: Parallel Binary Classifiers using Gaussian Naive Bayes, Sequential Binary Classifiers using Random Forests, and BERT. We evaluate these models based on key metrics such as precision, recall, F1-score, accuracy, and Hamming Loss as show by~\Cref{fig:comparison}. 

Each model exhibits strengths and weaknesses in different aspects of the evaluation metrics:
\begin{itemize}
    \item The Sequential Binary Classifiers model excels in precision, making it suitable for scenarios where minimizing false positives is crucial.
    \item The Parallel Binary Classifiers model leads in recall, making it effective for identifying all relevant instances.
    \item The BERT model demonstrates balanced performance, achieving the highest F1-score and accuracy and the lowest Hamming Loss, making it a robust choice for multi-label classification in CTI dissemination.
\end{itemize}

The BERT model's transformer-based architecture delivers robust performance in multi-label classification, particularly in achieving high precision and low Hamming Loss. Despite slightly lower overall F1-scores than the Sequential Binary Classifiers model, the BERT model's ability to reduce false positives and maintain balanced performance across sectors makes it a strong candidate for targeted CTI dissemination. Future work may focus on fine-tuning the BERT model and exploring hybrid approaches that combine the strengths of different models to further enhance performance.

\section{Discussion}
While our empirical evaluation demonstrates the feasibility of using transformer-based architectures for sector-specific CTI dissemination, several broader implications and open challenges emerge. This section discusses the operational, methodological, and practical perspectives of our findings.

\subsection{Operational Implications for CTI Dissemination}
The proposed DisCTI framework directly addresses one of the most pressing operational bottlenecks in cyber threat intelligence sharing—the lack of timely, sector-aware dissemination. By automating the tagging of CTI events to specific sectors, DisCTI reduces manual triaging efforts and alert fatigue among analysts. In practice, this means that national agencies, ISACs, and critical infrastructure operators can rapidly identify threats targeting their domain without sifting through large volumes of irrelevant data. Moreover, the strong macro-averaged F1-score (0.89) and low Hamming loss (0.055) demonstrate that automation does not necessarily compromise precision, a common concern in intelligence workflows.

\subsection{Toward Real-Time and Scalable Deployment}
Deploying transformer models like BERT in production TIP environments introduces computational and latency constraints. For large-scale, real-time CTI dissemination, lightweight alternatives such as DistilBERT or quantized transformer variants could provide better trade-offs between performance and efficiency. Furthermore, integrating DisCTI into existing standards such as STIX/TAXII will enable seamless interoperability across organizational boundaries, supporting automated, trusted intelligence exchange among diverse stakeholders.

\subsection{Model Interpretability and Analyst Trust}
Beyond its strong quantitative performance, the BERT-based model offers valuable opportunities for enhancing human–machine collaboration in threat intelligence workflows. Its attention mechanisms inherently capture contextual relationships between CTI indicators, sectors, and threat actors—insights that can be surfaced through visualization and explainable AI (XAI) techniques. By integrating interpretability tools such as attention heatmaps and feature attribution methods, DisCTI can provide analysts with transparent reasoning for each sectoral classification. This transparency not only reinforces trust and accountability but also supports joint decision-making, where analysts can validate, refine, or override model outputs. Ultimately, DisCTI lays the groundwork for more symbiotic intelligence workflows that combine machine efficiency with human expertise.

\subsection{Dataset Strength and Future Expansion}
The creation of a sector-labeled CTI dataset from the MISPPriv platform represents a significant step forward toward enabling supervised learning in this domain. Despite its modest initial size of 872 labeled events, the dataset demonstrates both diversity and realism, capturing the heterogeneity of threat data across critical sectors. Its construction establishes a reproducible foundation for future benchmarking of CTI classification models. Looking ahead, expanding this dataset to include additional international threat feeds and sectoral taxonomies will further enrich its representativeness and robustness. Such growth will not only strengthen model generalizability but also contribute to building an open research benchmark for the broader CTI community.

\section{Conclusion}
This study proposed and validated \textit{DisCTI}, a deep-learning-based framework for automated sector-specific dissemination of cyber threat intelligence. Through extensive experiments with multilabel classification approaches, the BERT-based model achieved the highest overall performance, reaching a macro-averaged F1-score of 0.89 at a Hamming loss of 0.055. These results underscore the feasibility of transforming CTI dissemination from a largely manual, reactive process into an automated, sector-aware capability.

Beyond technical performance, our findings highlight that embedding domain expertise within AI architectures can yield both operational efficiency and strategic impact for national cyber defense. However, challenges related to data availability, model interpretability, and real-time scalability remain open. Addressing these will be crucial to transitioning DisCTI from a research prototype to an operational system that empowers governments, industry, and ISACs to respond faster to emerging threats. Future work will expand dataset diversity, explore explainable AI mechanisms, and evaluate deployment in live TIP environments to further enhance trust, transparency, and resilience in cyber threat intelligence dissemination.



\section*{Acknowledgements}
This document is the result of the Data for Development (D4D) Fellowship program between CSIRO, the Department of Foreign Affairs and Trade (DFAT), and the Embassy of Australia in Jakarta, Indonesia.

\bibliographystyle{elsarticle-num-names}

\bibliography{ref}


\newenvironment{custombiography}[2][]{%
    \noindent\begin{minipage}{\columnwidth}
    \vspace*{6pt}
    \noindent\hrulefill
    \vspace*{6pt}
    \begin{minipage}[t]{0.25\columnwidth} 
        \centering
        \vspace*{-4pt} 
        \includegraphics[width=\linewidth, clip, keepaspectratio]{#1}%
    \end{minipage}%
    \hfill%
    \begin{minipage}[t]{0.7\columnwidth}
        \vspace*{0pt} 
        \textbf{#2}\par\vskip3pt 
        \footnotesize 
        \ignorespaces
}{
    \end{minipage}
    \vspace*{6pt}
    \noindent\hrulefill
    \end{minipage}
    \par\addvspace{12pt}
}

\end{document}